# Translationally deformed topological charge nanolaser with an ultrasmall mode volume

**Shengqun Guo[1], Feng Tian[1], Yuhua Liao[1], Bowen Han[1], and Taojie Zhou[1,*]**

[1] School of Microelectronics, South China University of Technology, Guangzhou, 511442, China
*corresponding author: Taojie Zhou (taojiezhou@scut.edu.cn)

Developing vortex nanolasers is highly desirable for on-chip multidimensional large-capacity information processing. Topological optical modes hold great promise for achieving coherent emission with diverse functionalities. However, the development of robust and ultracompact topological charge lasing operation remains insufficiently explored. Here, we theoretically propose a translationally deformed topological charge vortex nanocavity with a low mode volume of 0.32 $(\lambda/n)^3$, and experimentally demonstrate the corresponding lasing emission with a low lasing threshold of around 0.74 μW. The designed topological nanocavity, constructed by translationally deformed photonic crystals, supports an ultracompact optical mode carrying a topological charge characterized by polarization winding. The well-defined topological charge characteristics of the fabricated device are revealed in both near- and far-field polarization-resolved optical profiles. Our work opens a promising avenue for versatile topological photonic integration and gives new potential for exploring intriguing structured light-matter interactions under the topological photonics scenario.

**INTRODUCTION**

Over the past few decades, there has been a surge of substantial attention in the polarized optical vortices with a nonzero topological charge, due to their distinctive optical characteristics and broad applications in both fundamental physics and various applications[1-6]. Driven by the demand for compact on-chip solutions, notable advances in the development of optical vortices with topological charge created directly in micro/nano photonic crystal (PhC) structure have been reported based on the bound states in the continuum[7,8] and Dirac-vortex state[9-11]. Although impressive, the loose physical footprint and low modal confinement hamper their prospects for realizing ultra-density integrated on-chip light source with low-energy consumption. As such, pursuing ultracompact robust nanolasers with ultralow lasing thresholds capable of topological charge remains a significant challenge, primarily constrained by the fundamental physical design principle of optical vortex cavities. One of the successful efforts to address this issue is introducing the concept of topological crystal disclination utilizing complicated Volterra processing to construct the photonic disclination cavity[12,13], accompanied by further precise optimization of the tailored disclination core to achieve lasing operation.

In the broader arena of topological photonics, light propagation and localization under different classes of photonic topological insulators have been investigated extensively[14-18], catalyzing the development of the reliable on-chip light source with robustness against structural perturbations. Motivated by the topological ideas of quantum spin-Hall and valley Hall effects[19-24], pioneering topological lasers utilizing spin-Hall and valley-Hall type edge modes have been demonstrated[25-29], along with the realization of high-performance lasers based on the band-inversion-confined bulk state and valley PhC heterostructure[30-33]. Beyond the scope of conventional topological bulk and edge lasers with a relatively large mode volume ($V_m$) and weak light-matter interaction, recently reported higher-order topological corner state nanolasers enable light trapping from the high quality factor ($Q$-factor) zero-dimensional (0D) optical mode, which facilitates the achievement of ultralow threshold lasing emission in an ultracompact footprint[34-41]. Nevertheless, previously demonstrated single corner state topological nanolasers have been mainly limited to a specific Zak phase distinction PhC configuration and do not exhibit the lasing characteristics associated with polarized optical vortices. It thus remains challenging to develop a simple yet high-quality topological nanocavity with a wavelength-scale $V_m$ for ultralow threshold topological charge lasing emission.

In this work, we present a new class of topological charge nanolaser based on the straightforward translationally deformed PhC structures. To start with, we demonstrate topological PhCs by considering the translational deformations, revealing that a 0D optical vortex mode is tightly confined in a nanocavity constructed by splicing PhCs with distinct translation directions. The designed high $Q$-factor and small $V_m$ vortex mode is strongly localized around the structural core with quadrupolar modal distribution and capable of polarization winding with topological charge characteristics. Experimentally, using the InGaAsP multi-quantum wells as the gain materials, stable single-mode lasing emission at room temperature with a low threshold of ~ 0.74 μW is achieved, of which the corresponding topological charge features of the lasing mode are also uncovered. Our work provides a simple route toward ultracompact topological light sources with unique polarization properties and highlights promising directions to manipulate light and explore topological phenomena under translational deformation.

**RESULTS AND DISCUSSION**

Fig. 1(a) schematically shows the considered square-lattice PhC structure, in which the initial state hosts a square-shaped air hole with side length *s* at the center of the PhC unit cells (UCs), corresponding to two-dimensional $C_{4v}$ PhC under the trivial atomic limit[42,43]. Afterward, a translational deformation is introduced to shift the air hole away from the center positions (indicated by the dashed squares) in vertical or horizontal directions, thereby separating the UCs into four configurations, as shown in Fig. 1(a), where the periodic translational parameter *d* describes the relative shift ranging from −*a*/2 to *a*/2. Figure 1b displays the calculated transverse-electric (TE) like band structure for the initial state with parameters *a* = 485 nm, *s*/*a* = 0.6, thickness of slab *h* = 260 nm, and refractive index *n* = 3.25, revealing an omnidirectional bulk band gap from 0.31 *a*/*λ* to 0.34 *a*/*λ* between the first and second bands. For the configurations with varied *d*, they share the identical photonic band diagram since they belong to distinct options of UC for the same PhC structure. To characterize the topological feature of different configurations under translational deformation, Fig. 1(c)-(d) present the eigenmode profiles for the extreme case with translational parameter *d* = 0.5*a* in horizontal and vertical directions, respectively. For horizontal configuration, the parity of eigenmode profiles at the X and Y points of the occupied band is odd and even, while it exhibits opposite parities at the high-symmetry points in vertical configuration, indicating the topologically distinct feature under different translational directions. Due to the remaining reflection symmetry, the parity of eigenmode profiles can be reflected to quantized two-dimensional Zak phase[44-46]. For horizontal configuration, the value of the Zak phase is ($\pi$, 0), namely, the non-trivial bulk polarization $P_x = 1/2$ and trivial $P_y = 0$ under reduced crystalline symmetry. Similarly, one can obtain the value of the Zak phase as (0, $\pi$) for vertical configuration with $d = 0.5a$. These results indicate that half-period translational deformation generates a non-trivial Zak phase in a single direction, i.e., the PhC enters into obstructed atomic limit topological phases, wherein the Wannier centers are located at the Wyckoff position of the unit's edges[42,43]. Although the non-trivial single direction Zak phase typically reflects the emergence of only horizontal or vertical topological edge states, the polarization vector is not equivalent with respect to the horizontal and vertical translation directions, thus providing the possibilities for 0D confined states in the finite-size domain with topologically distinct surroundings.

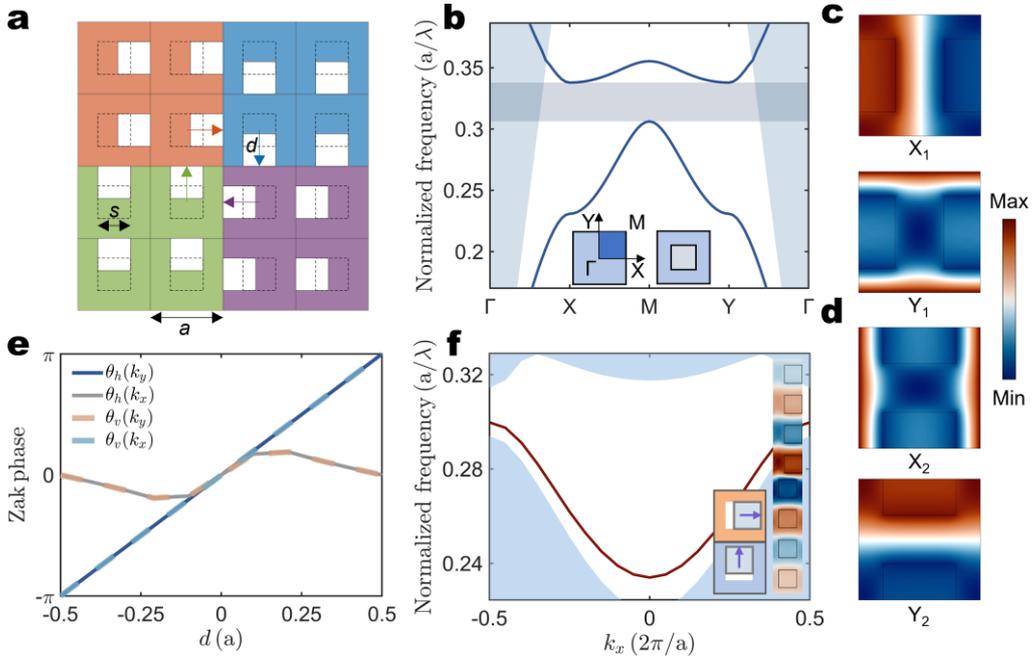

**FIG. 1. Photonic crystal under translation deformation strategy.** (a) Schematic diagram of the distinct translation deformation path in PhCs. The dashed and solid squares represent air holes before and after the translation deformation, respectively. The arrows illustrate the shift directions. (b) The TE-like photonic band structure with light line and bandgap decorated in blue and gray regions. The insets show the first Brillouin zone and representative UC. (c, d) The eigenmode profiles of the first bulk band at high-symmetry points for the extreme case under horizontal (c) and vertical (d) directions. (e) The evolution of the Zak phase as a function of the translational parameter *d*. (f) Projected band structures of the supercell (inset). The edge state is labeled as the red line.

For more general translational deformation configurations with $d \neq 0$, the vanishing of the centered reflection axis (i.e., UC reflection symmetry) implies that the Zak phase will not necessarily be $\pi$ quantized[47,48]. Figure 1e shows the evolution of Zak phases with scanning $d$ in a period by using the Wilson loop approach[49-51]. The Zak phase $\theta_h(k_y)$ of the occupied bulk band for horizontal translation configurations is continuous linear evolution from 0 to $2\pi$, while $\theta_h(k_x)$ does not wind by progressively changing the translation parameter. Moreover, the winding of Zak phase implies the existence of a non-zero integer Chern number that is independent of structural details[51-53], in which the situation is quite different for horizontal and vertical configurations. The $2\pi$ phase winding of $\theta_h(k_y)$ for horizontal translation configuration results in a Chern number $C(k_y) = 1$, while the $\theta_v(k_y)$ remains trivial with $\theta_v(k_x)$ evolving $2\pi$ for vertical translation configuration corresponding to Chern numbers $C(k_y) = 0$ and $C(k_x) = 1$. The topological distinction for different translation directions enables 1D edge states along the crystalline interfaces between the two PhC regions. To elucidate it, the horizontal and vertical translations are applied to the top and bottom halves of the PhC divided along the $y$ direction, forming a crystalline interface between the two PhC regions where an edge state around the interface is observed within the bandgap in the projected band structure shown in Fig. 1(f).

To illustrate the existence of the 0D confined mode in the topological PhC nanocavity with translational deformation, Fig. 2(a) shows the schematic of the proposed topological charge nanocavity, which is constructed by splicing four quadrants with distinct translational directions indicated by different color patches, forming a structural core located at the intersection of the deformed interfaces. Figure 2b displays the eigenfrequency spectra of this architecture with $a = 485$ nm, $s/a = 0.6$, and $d = 90$ nm, clearly revealing an in-gap single 0D confined mode (indicated in pentagram) around 196 Thz. This confined mode hosts a high *Q*-factor of approximately $1.7\times10^4$ and an ultrasmall wavelength-scale $V_m$ around 0.32 $(\lambda/n)^3$, contributing to a high Purcell factor and thus facilitating the attainment of the ultralow threshold topological lasing emission from the confined mode. To further characterize the 0D confined mode, the corresponding simulated $H_z$ field distribution is present in Fig. 2(c). The confined mode is tightly localized around the structural core and hosts a quadrupolar modal distribution profile, providing the opportunity for vortex emission with unique polarization characteristics. Furthermore, the phase distributions of ($E_x$ + i$E_y$) shown in Fig. 2(d) reveal that the confined mode hosts a winding of the polarization characterized by a closed loop around the singularity, signaling a nontrivial topological charge of |*q*| = 1 for the deformed nanocavity. To demonstrate the polarization characteristics of the proposed nanocavity, Figs. 2(e)-(g) present total and polarization-resolved Poynting vector distributions. It can be seen that the total Poynting vector distributions in Fig. 2(e) show a doughnut-like pattern with a central dark node. For the polarization-resolved Poynting vector distributions shown in Figs. 2(f)-(g), the two bright lobes rotate by 90 degrees depending on polarization direction and are parallel to the white arrow, indicating the nature of radial polarization. These features distinguish the proposed translationally deformed nanocavity from previously demonstrated topological corner state cavity without the polarized optical vortices characteristics[34-41].

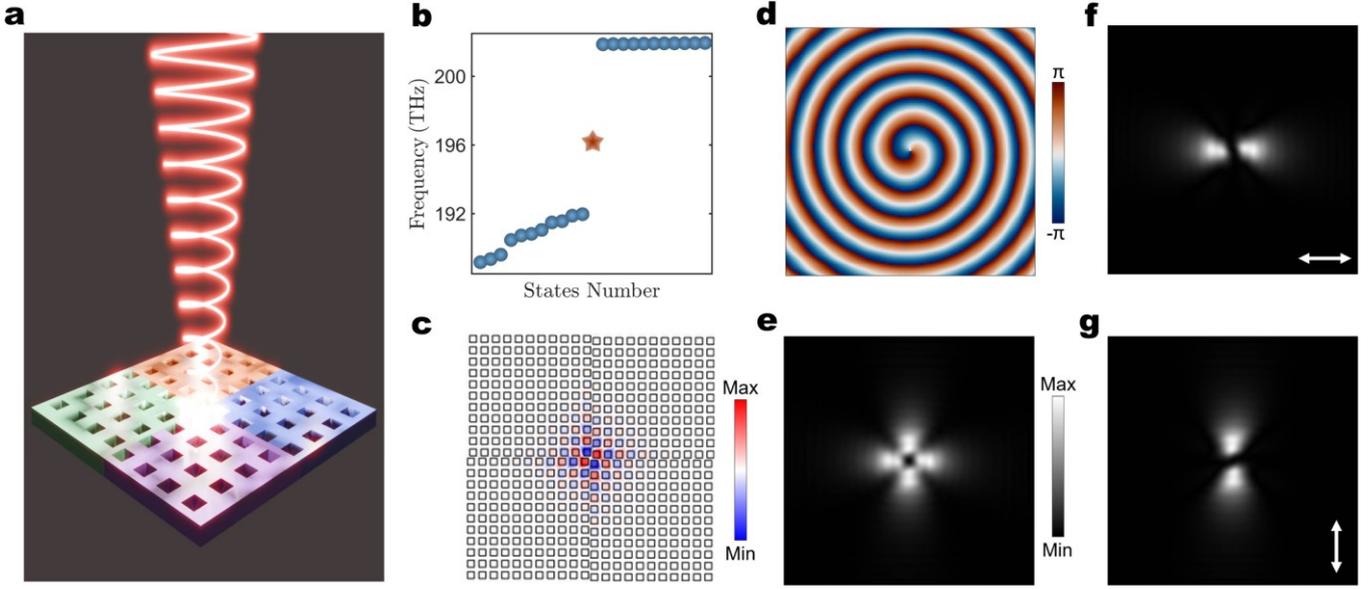

**FIG. 2. Topological nanocavity design. (a)** Conceptual illustration of the proposed translationally deformed topological charge nanocavity. **(b)** The eigenfrequency distribution for nanocavity following the design in **(a)**. **(c, d)** The simulated field profiles $H_z$ **(c)** and phase distribution **(d)** for the confined mode. **(e-g)** The simulated total **(e)**, horizontally polarized **(f)**, and vertically polarized **(g)** Poynting vectors distribution above the slab.

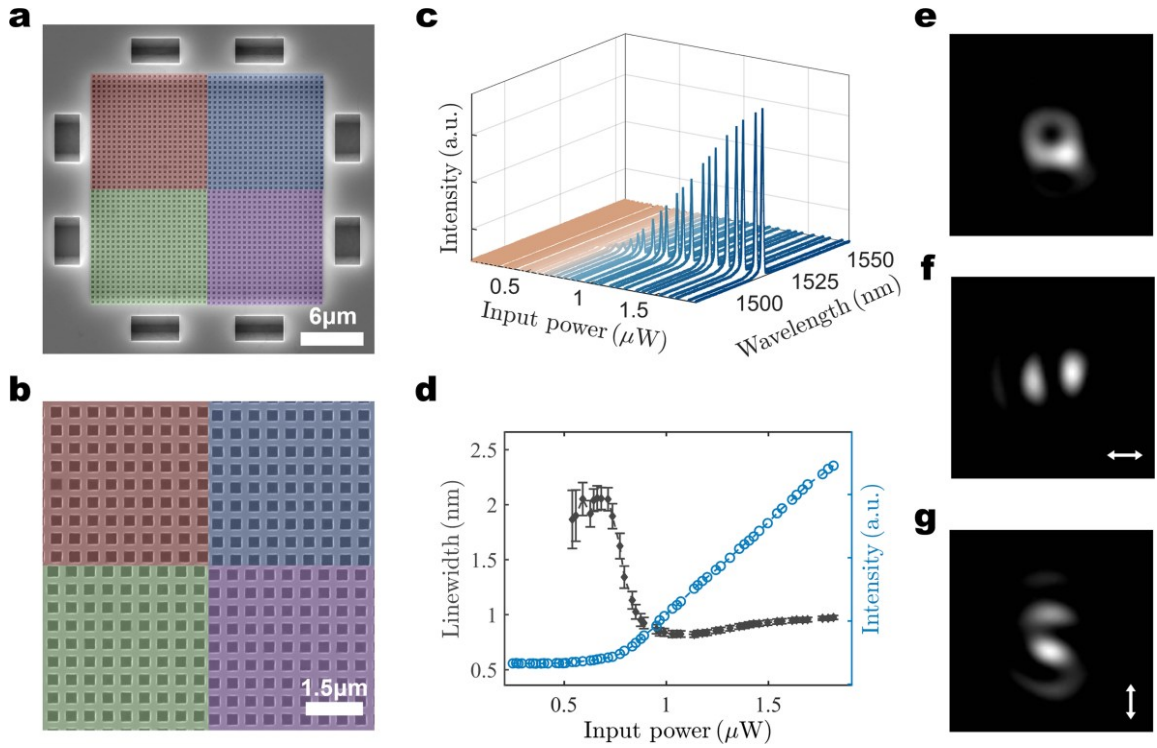

**FIG. 3 Experimental characterizations of the fabricated topological charge nanolaser. (a, b)** The top-view of the full PhC pattern **(a)** and the zoomed-in **(b)** SEM image of the fabricated device. **(c)** Emission spectra evolution under various input powers. **(d)** *L-L* curve (blue) and the evolution of linewidth (black) of the nanolaser as a function of the average input power. **(e-g)** The captured lasing mode image without polarizer **(e)** and the polarization-resolved images with horizontal **(f)** and vertical **(g)** polarization components, respectively.

To experimentally realize the associated vortex lasing emission, the device is fabricated on a 260 nm-thick InGaAsP multiple quantum wells platform using standard semiconductor nanofabrication processes[40,41,54]. Figures 3a-b present the top view of the full PhC pattern and the associated enlarged core region for a fabricated device captured by scanning electron microscopy (SEM). Each quadrant consists of 23 periods of UCs, and the lateral size of the full device is ~ 22 microns with the parameters nominally identical to the simulation. To assess the optical characteristics of the fabricated translationally deformed topological charge nanolasers, room temperature micro-photoluminescence (μ-PL) measurement of the device was performed with a 632 nm pulsed laser (200 kHz repetition rate and 0.5% duty cycle). The power-dependent emission spectra shown in Fig. 3(c) demonstrate a clear spectral evolution from spontaneous emission dominance at lower input power to single-mode lasing action as input power increases. Such single-mode lasing action, identified by the narrow lasing peak, emerges around 1515 nm and operates over a wide pumping range. Figure 3d presents its light in-light out (*L-L*) curve as well as the linewidth evolution to further quantify the occurrence of coherent emission, which exhibits the clear lasing behavior that includes the distinct threshold kink and the linewidth narrowing phenomenon. An ultralow lasing threshold of ~ 0.74 μW (power density 4.71 kW/cm$^2$) is achieved, mainly attributed to the enhanced light-matter interaction in the proposed topological charge nanocavity with a high *Q*-factor and a low $V_m$. To further verify the lasing mode, the polarization-resolved near-field lasing profiles are captured by an infrared camera as shown in Figs. 3(e)-(g). The lasing mode pumped from the core region without a linear polarizer shows a doughnut-like profile, which matches well the simulated result shown in Fig. 2(e). The mode profiles under the polarizer in front of the camera [Figs. 3 (f)-(g)] also agree well with the polarization-resolved Poynting vector distributions shown in Figs. 2(f)-(g), exhibiting the typical two-lobe-structured profiles and directly verifying that the single-mode lasing originated from the predicted confined mode. Moreover, Figs. 4(a)-(b) present the experimental and simulated far-field emission patterns. The experimental result without the polarizer, as shown in the first panel of Fig. 4(a), presents a $C_4$ symmetric beam profile with a central intensity node. As expected for the confined topological charge vortex mode, the corresponding *x* and *y* components of the far-field pattern present the polarization-resolved beam profiles with two lobes, which match well with the simulated ones shown in Fig. 4(b). Collectively, these results indicate that the lasing emission is attributed to the predicted 0D confined mode and further corroborate the topological strategy to broaden the versatility of nanolasers.

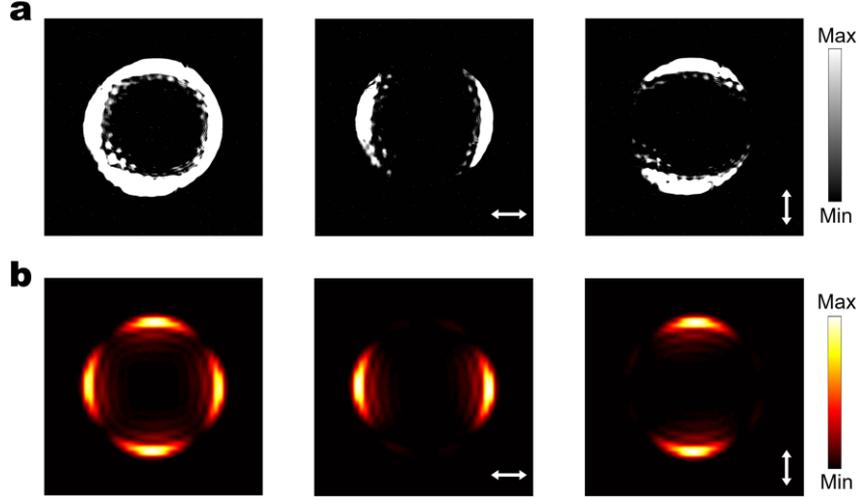

**FIG. 4 Experimental and simulated far-field results. (a)** The measured far-field emission patterns from the designed nanolaser. A polarizer was rotated in front of the infrared camera to resolve the polarization-resolved profiles. **(b)** Simulated far-field and polarization-resolved far-field patterns.

## CONCLUSION

In conclusion, we present the topological charge lasing emission based on a 0D confined mode in a new class of topological PhC nanocavity, which is spliced by translationally deformed PhCs with abundant topological features. To confirm topological charge lasing action, the fabricated nanolaser was optically pumped at room temperature and characterized through μ-PL measurement and polarization-resolved optical characterizations. The strong light confinement, characterized by a high $Q$ and small $V_m$ for the designed topological nanocavity, leads to single-mode lasing operation with an ultralow lasing threshold of ~ 0.74 μW. The polarization-resolved near-field and far-field optical patterns further reveal the polarization characteristics of the topological charge lasing, which match well with simulation results. Our work reveals a wavelength-scale confined mode in translation deformed topological nanocavity with unique polarization characterized, representing a promising platform for pursuing robust nanoscale on-chip light sources capable of topological charge. In parallel, this demonstration provides a feasible strategy for polarization-dependent confined states in the context of topological photonics and is attractive for studying the interaction between structured light and matter. Considering the fact that the topology concepts are relevant in various artificial bandgap materials[55-57], our design is not limited to nanolasers but also may prompt further topological research in PhC fibers[58,59] and so on.

## MATERIALS AND METHODS

### Device fabrication

The translationally deformed PhC nanolaser was fabricated based on the InP epitaxial wafer with InGaAsP multiple quantum wells as the gain materials. A 90-nm-thick $SiO_2$ hard mask was deposited using plasma-enhanced chemical vapor deposition, and electron-beam lithography was subsequently employed to define the PhC pattern in the PMMA electron resist layer. Using the inductively coupled plasma dry etching method, the designed pattern is further transferred into the gain layer. Afterwards, the residual PMMA layer and $SiO_2$ hard mask were removed by $O_2$ plasma and buffered oxide etching solution, respectively. Finally, an InP sacrificial layer beneath the gain slab was wet-etched using a diluted hydrochloric acid solution to form the final suspended PhC structure.

### Optical measurement

The fabricated devices were characterized using the μ-PL system with a nanosecond 632 nm pulsed laser employed as the excitation source. A 100× objective lens with a high numerical aperture (0.9) was used to focus the pumping beam and collect the emission from the device. The emission spectra were analyzed by the spectrometer with a liquid-nitrogen-cooled infrared InGaAs detector. The near-field images were captured by an InGaAs camera with a linear polarizer employed to measure the polarization-resolved patterns. 4f optical system is used to obtained the far-field patterns.

**Numerical simulations**

In numerical studies, the photonic band structure, eigen-frequencies, $Q$-factor, $V_m$, Poynting vector distributions, and far-field patterns were calculated by the finite-element method. The refractive indices of the dielectric slab and air were assumed to be 3.25 and 1, respectively. The $V_m$ is calculated referring to $V_m = \frac{\int \varepsilon(\mathbf{r})|E(\mathbf{r})|^2 dV}{max(\varepsilon(\mathbf{r})|E(\mathbf{r})|^2)}$, where $\varepsilon(\mathbf{r})$ and $E(\mathbf{r})$ are the spatially dependent dielectric constant and electric field intensity, respectively.

**Acknowledgements:** T.Z. acknowledges the financial support from the National Natural Science Foundation of China (Grant No. 62304080), the Guangdong Basic and Applied Basic Research Foundation (Grant No. 2024A1515010802), the Science and Technology Projects in Guangzhou (Grant No. 2024A04J3683), the Fundamental Research Funds for the Central Universities (Grant No. 2023ZYGXZR068), and the startup funds from South China University of Technology. The authors acknowledge the support from the Micro & Nano Electronics Platform (MNEP) of SCUT for device fabrication and characterization, and the Micro/Nano Fabrication Platform of the Institute of Semiconductors, Guangdong Academy of Sciences, for support in device fabrication. **Author contributions:** S.G. and T.Z. conceived the idea. S.G. carried out the numerical calculations and simulations. F.T and Y.L. fabricated the devices. S.G. performed the optical measurements with assistant from B.H.. S.G. analyzed the data and wrote the manuscript with inputs from all co-authors. T.Z. supervised the research project. **Competing interests:** The authors declare no competing interests. **Data and materials availability:** The data that support the findings of this study are available from the corresponding author upon reasonable request.